\documentstyle[12pt]{article}
\input epsf

\begin{document}
\baselineskip=20.5pt

\def\beqra{\begin{eqnarray}} \def\eeqra{\end{eqnarray}}
\def\beqast{\begin{eqnarray*}} \def\eeqast{\end{eqnarray*}}
\def\beq{\begin{equation}}      \def\eeq{\end{equation}}
\def\be{\begin{enumerate}}   \def\ee{\end{enumerate}}

%title page
\def\fnote#1#2{\begingroup\def\thefootnote{#1}\footnote{#2}\addtocounter
{footnote}{-1}\endgroup}

\def\ut#1#2{\hfill{UTTG-{#1}-{#2}}}
\def\fl#1#2{\hfill{FERMILAB-PUB-94/{#1}-{#2}}}
\def\itp#1#2{\hfill{NSF-ITP-{#1}-{#2}}}

\def\bet{\beta}
\def\gam{\gamma}
\def\Gam{\Gamma}
\def\la{\lambda}
\def\eps{\epsilon}
\def\La{\Lambda}
\def\si{\sigma}
\def\Si{\Sigma}
\def\al{\alpha}
\def\Tha{\Theta}
\def\tha{\theta}
\def\vphi{\varphi}
\def\del{\delta}
\def\Del{\Delta}
\def\ab{\alpha\beta}
\def\om{\omega}
\def\Om{\Omega}
\def\mn{\mu\nu}
\def\mun{^{\mu}{}_{\nu}}
\def\kap{\kappa}
\def\rsi{\rho\sigma}
\def\beal{\beta\alpha}

\def\til{\tilde}
\def\rta{\rightarrow}
\def\eqv{\equiv}
\def\nab{\nabla}
\def\pa{\partial}
\def\sit{\tilde\sigma}
\def\ul{\underline}
\def\indt{\parindent2.5em}
\def\nd{\noindent}

\def\rsi{\rho\sigma}
\def\beal{\beta\alpha}

        % calligraphic
\def\caa{{\cal A}}
\def\cb{{\cal B}}
\def\cac{{\cal C}}
\def\cd{{\cal D}}
\def\ce{{\cal E}}
\def\cf{{\cal F}}
\def\cg{{\cal G}}
\def\ch{{\cal H}}
\def\ci{{\cal I}}
\def\cj{{\cal{J}}}
\def\ck{{\cal K}}
\def\cl{{\cal L}}
\def\cm{{\cal M}}
\def\cn{{\cal N}}
\def\cO{{\cal O}}
\def\cp{{\cal P}}
\def\car{{\cal R}}
\def\cs{{\cal S}}
\def\ct{{\cal{T}}}
\def\cu{{\cal{U}}}
\def\cv{{\cal{V}}}
\def\cw{{\cal{W}}}
\def\cx{{\cal{X}}}
\def\cy{{\cal{Y}}}
\def\cz{{\cal{Z}}}

\def\cdgc{C^{\dagger}C}
\def\ccdg{CC^{\dagger}}
\def\pcdgc{C'^{\dagger}C'}
\def\pccdg{C'C'^{\dagger}}
\def\vdgv{v^{\dagger}v}
\def\gmn{\cg^{\mu}_\nu}
\def\gmm{\cg^{\mu}_\mu}
\def\gmnb{\cg^{~\,\mu}_{0\,\nu}}
\def\smn{\Sigma^{\mu}_\nu}
\def\pdgp{\phi^{\dagger}\phi}
\def\ga{\cg^{(a)}}
\def\gamn{\cg^{(a)\mu}_{~~\nu}}
\def\gamm{\cg^{(a)\mu}_{~~\mu}}
\def\sia{\Si^{(a)}}
\def\siamn{\Si^{(a)\mu}_{~~\nu}}
\def\Gak{\Gam^{(a)}_{2k}}
\def\invp{\cg_0^{-1}}

        % nots
\def\raisenot{\raise .5mm\hbox{/}}
\def\nota{\ \hbox{{$a$}\kern-.49em\hbox{/}}}
\def\notA{\hbox{{$A$}\kern-.54em\hbox{\raisenot}}}
\def\notb{\ \hbox{{$b$}\kern-.47em\hbox{/}}}
\def\notB{\ \hbox{{$B$}\kern-.60em\hbox{\raisenot}}}
\def\notc{\ \hbox{{$c$}\kern-.45em\hbox{/}}}
\def\notd{\ \hbox{{$d$}\kern-.53em\hbox{/}}}
\def\notbd{\ \hbox{{$D$}\kern-.61em\hbox{\raisenot}}} %big D
\def\note{\ \hbox{{$e$}\kern-.47em\hbox{/}}}
\def\notk{\ \hbox{{$k$}\kern-.51em\hbox{/}}}
\def\notp{\ \hbox{{$p$}\kern-.43em\hbox{/}}}
\def\notq{\ \hbox{{$q$}\kern-.47em\hbox{/}}}
\def\notW{\ \hbox{{$W$}\kern-.75em\hbox{\raisenot}}}
\def\notz{\ \hbox{{$Z$}\kern-.61em\hbox{\raisenot}}}
\def\notpa{\hbox{{$\partial$}\kern-.54em\hbox{\raisenot}}}

\def\fo{\hbox{{1}\kern-.25em\hbox{l}}}  %raised one
\def\rf#1{$^{#1}$}
\def\bx{\Box}
\def\tr{{\rm Tr}}
\def\rmtr{{\rm tr}}
\def\dgg{\dagger}
\def\trm{{\rm tr}_{_{\left(M\right)}}~}
\def\trn{{\rm tr}_{_{\left(N\right)}}~}
\def\trnn{{\rm tr}_{_{\left(N+1\right)}}~}
\def\tr2n{{\rm tr}_{_{\left(2N\right)}}~}

\def\lag{\langle}
\def\rag{\rangle}
\def\bmid{\big|}

\def\vlap{\overrightarrow{\La p}} %overrightarrow or left will grow over 
\def\lrta{\longrightarrow} \def\lrar{\raisebox{.8ex}{$\longrightarrow$}}
\def\rlarw{\longleftarrow\!\!\!\!\!\!\!\!\!\!\!\lrar}
\def\vx{\vec x}  % x vector
\def\vy{\vec y}  % y vector

\def\llra{\relbar\joinrel\longrightarrow}           %THIS IS LONG
\def\mapright#1{\smash{\mathop{\llra}\limits_{#1}}} %ARROW ON LINE
\def\mapup#1{\smash{\mathop{\llra}\limits^{#1}}} %CAN PUT SOMETHING OVER 

\def\nmasymptotic{
{_{\displaystyle{\rm lim}}\atop
{\scriptstyle N,M\rightarrow\infty}
}\,\, 
}

\def\nasymptotic{{_{\stackrel{\displaystyle\longrightarrow}
{N\rightarrow\infty}}\,\, }} %N  goe to infinity, display sty.
\def\masymptotic{{_{\stackrel{\displaystyle\longrightarrow}
{M\rightarrow\infty}}\,\, }} %M  goe to infinity, display sty.
\def\wasymptotic{{_{\stackrel{\displaystyle\longrightarrow}
{w\rightarrow\infty}}\,\, }} %w goe to infinity, display sty.

\def\asymptext{\raisebox{.6ex}{${_{\stackrel{\displaystyle\longrightarro
w}{x\rightarrow\pm\infty}}\,\, }$}} %x goes to plus minus infinity, within 
\def\epsilim{{_{\textstyle{\rm lim}}\atop_{\epsilon\rightarrow 0+}\,\, }} %epsilon goes to zero, display 

\def\7#1#2{\mathop{\null#2}\limits^{#1}}        % puts #1 atop #2
\def\5#1#2{\mathop{\null#2}\limits_{#1}}        % puts #1 beneath #2
\def\too#1{\stackrel{#1}{\to}}
\def\tooo#1{\stackrel{#1}{\longleftarrow}}
\def\nout{{\rm in \atop out}}

\def\one{\raisebox{.5ex}{1}}
\def\BM#1{\mbox{\boldmath{$#1$}}}

\def\ltsim{\matrix{<\cr\noalign{\vskip-7pt}\sim\cr}}
\def\gtsim{\matrix{>\cr\noalign{\vskip-7pt}\sim\cr}}
\def\haf{\frac{1}{2}}

%       pictures

\def\place#1#2#3{\vbox to0pt{\kern-\parskip\kern-7pt
                             \kern-#2truein\hbox{\kern#1truein #3}
                             \vss}\nointerlineskip}

\def\illustration #1 by #2 (#3){\vbox to #2{\hrule width #1 height 0pt 
depth
0pt
                                       \vfill\special{illustration #3}}}

\def\scaledillustration #1 by #2 (#3 scaled #4){{\dimen0=#1 \dimen1=#2
           \divide\dimen0 by 1000 \multiply\dimen0 by #4
            \divide\dimen1 by 1000 \multiply\dimen1 by #4
            \illustration \dimen0 by \dimen1 (#3 scaled #4)}}

\def\ON{{\cal O}(N)}
\def\UN{{\cal U}(N)}
\def\bdPh{\mbox{\boldmath{$\dot{\!\Phi}$}}}
\def\bPh{\mbox{\boldmath{$\Phi$}}}
\def\bPhs{\bPh^2}
\def\sef{S_{eff}[\sigma,\pi]}
\def\sigx{\sigma(x)}
\def\pix{\pi(x)}
\def\bph{\mbox{\boldmath{$\phi$}}}
\def\bphs{\bph^2}
\def\ex{\BM{x}}
\def\exs{\ex^2}
\def\xdot{\dot{\!\ex}}
\def\y{\BM{y}}
\def\ys{\y^2}
\def\ydot{\dot{\!\y}}
\def\pat{\pa_t}
\def\pax{\pa_x}
\def\cia{C_{i\alpha}}
\def\cjb{C_{j\beta}}
\def\Gz{G(z)}
\def\log{{\rm log}~}
\def\Re{{\rm Re}~}
\def\Im{{\rm Im}~}
\def\nh{{\rm non-hermitean matrix}}
\def\det{{\rm det}~}
\renewcommand{\thesection}{\arabic{section}}
\renewcommand{\theequation}{\thesection.\arabic{equation}}

\itp{97}{025}
\today

\hfill{cond-mat/9704191}\\

\vspace*{.2in}

\begin{center}
{\large\bf NON-GAUSSIAN NON-HERMITEAN RANDOM MATRIX THEORY:\\
phase transition and addition formalism}\end{center}

\begin{center}
{\bf Joshua Feinberg$^{a *}$
 \& A. Zee$^{a,b}$\fnote{*}{{\it e-mail: joshua, zee@itp.ucsb.edu}}}

\end{center}
\vskip 2mm
\begin{center}
$^{a)}${Institute for Theoretical Physics}\\
{University of California\\ Santa Barbara, CA 93106, USA}\\
\vskip 2mm
$^{b)}${Institute for Advanced Study}\\
{Princeton, NJ 08540, USA}\\
\end{center}
\vskip 3mm
\begin{abstract}
We apply the recently introduced method of hermitization to study 
in the large $N$ limit non-hermitean random matrices that are drawn from a 
large class of circularly symmetric non-Gaussian probability distributions, 
thus extending the recent Gaussian non-hermitean literature.
We develop the general formalism for calculating the Green's function and averaged density of eigenvalues, which may be thought of as the non-hermitean analog of the method due to Br\`ezin, Itzykson, Parisi and Zuber for analyzing hermitean non-Gaussian random matrices. We obtain an explicit algebraic 
equation for the integrated density of eigenvalues. A somewhat surprising 
result of that equation is that the shape of the eigenvalue distribution in the complex plane is either a disk or an annulus. As a concrete example, we analyze the quartic ensemble and study the phase transition from a disk shaped eigenvalue distribution to an annular distribution. Finally, we apply the method of hermitization to develop the addition formalism for free non-hermitean random variables. We use this formalism to state and prove a non-abelian non-hermitean version of the central limit theorem. 
\end{abstract}

\vspace{25pt}
%PACS numbers: 11.10.Lm, 11.15.Pg, 11.10.Kk, 71.27.+a  
\vfill
\pagebreak

\setcounter{page}{1}

\section{Introduction}
\setcounter{equation}{0}

There has been considerable interest in random non-hermitean matrices recently. 
Possible applications range over several areas of physics\cite{nonherm, stony}. One difficulty is that the eigenvalues of non-hermitean matrices invade the complex plane, and consequently, various methods developed over the years to deal with random hermitean matrices are no longer applicable, as these methods typically all involve exploiting the powerful constraints of analytic function theory. (See in particular the paper by Br\'ezin, Itzykson, Parisi, and Zuber\cite{BIPZ}.) In a recent paper\cite{fz}, we proposed a ``method of 
hermitization", whereby a problem involving random non-hermitean matrices can 
be reduced to a problem involving random hermitean matrices, to which various 
standard methods (such as the diagrammatic method\cite{bzw}, or the ``renormalization group" method\cite{french, rg, daz, rectangles}) can be applied.

To our knowledge, the literature on random non-hermitean 
matrices\cite{nonherm, stony} has focussed 
exclusively on Gaussian randomness. For instance, it has been known for over thirty years, from the work of Ginibre\cite{ginibre}, that for the Gaussian probability distribution $P(\phi) =(1/Z) {\rm exp}~(-N\rmtr\phi^\dgg\phi)$ (here, as in the rest of this paper, $\phi$ denotes an $N\times N$ random matrix 
with the limit $N\rightarrow\infty$ understood), the density of eigenvalues of $\phi$ is 
uniformly distributed over a disk of radius 1 in the complex plane. In this 
paper, we point out that using the method of hermitization we can determine the 
density of eigenvalues of probability distribution of the form 
\beq\label{prob}
P(\phi) = {1\over Z} e^{-N\rmtr V(\phi^\dgg\phi)}\,,
\eeq
where $V$ is an arbitrary polynomial of its argument. Indeed, by a simple trick, we show that we can obtain the desired density of eigenvalues with a minimal amount of work, by judiciously exploiting the existing literature on random hermitean matrices. 

Due to the symmetry of $P(\phi)$ under the transformation $ \phi\rightarrow e^{i\alpha}\phi$, the density of eigenvalues is obviously rotational invariant. We find that the class of probability distributions of the form (\ref{prob}) exhibits a universal behavior in the sense that whatever
the polynomial $V$ is, the shape of the eigenvalue distribution in the complex plane is always either a disk or an annulus.

In a certain sense, our work may be thought of as the analog of the work of 
Br\'ezin et al. for random hermitean matrices \cite{BIPZ}; they showed how the density of eigenvalues of hermitean matrices $\varphi$ taken from the probability distribution $P(\varphi) = (1/Z){\rm exp} [-N\rmtr V(\varphi)]$ 
with $V$ an arbitrary polynomial can be determined, and not 
just for the Gaussian case studied by Wigner and others\cite{wigner}, in which $V=(1/2)\rmtr\varphi^2$. An important simplifying feature of the analysis in \cite{BIPZ} is that $P(\varphi)$ depends only on the eigenvalues of $\varphi$,
and not on the unitary matrix that diagonalizes it. In contrast, the probability distribution (\ref{prob}) for non-hermitean matrices depends explicitly on the $GL(N)$ matrix $S$ used to diagonalize $\phi=S^{-1} \Lambda S$, and $S$ does not decouple. Remarkably however, for the Gaussian $P(\phi)$, Ginibre \cite{ginibre} managed to integrate over $S$ explicitly and derived an explicit expression for the probability distribution of the eigenvalues of $\phi$. Unfortunately, it is not clear how to integrate over $S$ and derive the expression for the 
eigenvalue probability distribution for non-Gaussian distributions of the form (\ref{prob}). In this paper we circumvent this difficulty by using the method 
of hermitization.

As an explicit application of our method, we study the case $V(\pdgp)=m^2\pdgp + {g\over 2} (\pdgp)^2$ in detail. For $m^2$ positive, we expect we would get a disk-like distribution generalizing Ginibre's work. As we make $m^2\equiv -\mu^2$ more and more negative, we expect a phase transition at some critical value $\mu_c^2$, after which we might imagine the disk fragmenting into an annulus. Indeed, we find the critical value
\beqast
\mu_c^2 = \sqrt{2g}\,.
\eeqast 
We also calculate the density of eigenvalues inside the annulus in detail.

The problem of adding random hermitean matrices has been much discussed in the  
recent literature\cite{pastur, bzw, blue, bluez, bhz, zahed}. In particular, a Feynman diagrammatic proof of the formalism, written in terms of the so-called Blue's function, was given in \cite{bluez}. In a series of very interesting papers\cite{stony}, Zahed and his 
collaborators have extended the addition formalism to random non-hermitean 
matrices. It turns out that the formalism we used above to determine the density of eigenvalues of random non-hermitean matrices can be naturally used to study 
the addition formalism. Thus, we obtain, in a way which we naturally 
think is quite transparent, the addition formalism for random non-hermitean 
matrices, thus reproducing the result of Zahed et al. We use this addition formalism to state and prove a non-abelian non-hermitean version of
the central limit theorem.

\pagebreak
\section{The Two Dimensional Gas of Eigenvalues}
\setcounter{equation}{0}

As is well known, the Dyson gas approach represents one of the most
powerful methods in analyzing hermitean random matrices. Diagonalizing a
hermitean matrix $\phi=S^{\dagger}\Lambda S$ one writes
\beq\label{herjacobian}
\int d\phi f(\phi) = \int_{U(N)} d\mu(S) \int \prod_k d\lambda_k \left(\prod_{i<j} (\lambda_i - \lambda_j)^2\right)~ f(\lambda_k, S)\,.
\eeq
For non-hermitean random matrices, one can diagonalize $\phi=S^{-1}\Lambda
S$ and write 
\beq\label{complexjacobian}
\int d\phi d\phi^{\dagger} f(\phi, \phi^\dgg) = \int_{GL(N)} d\mu (S) \prod_k d\lambda_k d\lambda_k^*
\left(\prod_{i<j} |\lambda_i - \lambda_j|^4 \right)~f(\lambda_k,\lambda_k^*, S)\,.
\eeq
The fourth power in $(\lambda_i - \lambda_j)$ here, in contrast to the second 
power in (\ref{herjacobian}), is easily understood by dimensional analysis: a non-hermitean complex matrix has 
twice as many real variables as a hermitean matrix. The important difference between the hermitean and the non-hermitean cases is already 
apparent in the Gaussian case considered by Ginibre. The probability distribution 
\beq\label{gaussginibre}
P(\pdgp)={1\over Z}e^{-N \rmtr~ \phi^{\dagger}\phi}={1\over Z}e^{-N \rmtr
~[SS^{\dagger}\Lambda^* (SS^{\dagger})^{-1}\Lambda]}
\eeq
depends on $S$.
In the case of hermitean matrices, $S$ would be unitary and hence disappear from (\ref{gaussginibre}) entirely. Here, however, one must integrate over $S$. After some work,
Ginibre \cite{ginibre}
showed that
\beq\label{ginibreiz}
\int dS e^{-N ~\rmtr
[SS^{\dagger}\Lambda^* (SS^{\dagger})^{-1}\Lambda]}={\rm
constant} \prod_{i<j} |\lambda_i - \lambda_j|^{-2} e^{-N \sum_k
|\lambda_k|^2}
\eeq
and was thus able to proceed with the gas approach. Once
again, the problem reduces to the statistical mechanics of a
gas, this time in two dimensions, with a logarithmic repulsion between the 
gas molecules and confined by a harmonic potential.
Following Br\'ezin et al. \cite{BIPZ} in evaluating the integral by steepest descent, we obtain 
\beqast
\vx=\int d^2 y
~\rho(\vy) {\vx-\vy\over |\vx-\vy|^2}\,.
\eeqast
Making an analogy with two dimensional electrostatics mentioned, the 
solution of this equation is immediate: $\rho$ is equal to 
$1/\pi$ inside a disk of radius 1 centered at the origin, and vanishes
outside the disk.

We see however, that had we been faced with a more complicated probability
distribution, of the form $ P(\phi)={1\over Z}e^{-N ~\rmtr~
V(\phi^{\dagger}\phi)}={1\over Z}e^{-N~ \rmtr~
V[SS^{\dagger}\Lambda^*(SS^{\dagger})^{-1}\Lambda]}$, we do not
know how to carry out the integration over $S$. In this paper we circumvent 
this difficulty and present an alternative way of calculating the density of eigenvalues of $\phi$, based on the method of hermitization. It is an interesting challenge to obtain our results presented below using the 
gas method, and we conjecture to that end, that in the non-Gaussian case there 
should be a closed form expression for the integral over $S$ that is analogous to (\ref{ginibreiz}).

\pagebreak

\section{The Method of Hermitization}
\setcounter{equation}{0}
Here we briefly review the discussion in our recent paper\cite{fz}.
The averaged density of eigenvalues 
\beq\label{rho1}
\rho (x,y) =\langle {1\over N} \sum_i \delta(x-\Re \lambda_i)~ \delta (y- \Im
\lambda_i)\rangle
\eeq
of the non-hermitean matrix $\phi$, may be determined from the following two alternative quantities. The first quantity is 
\beq\label{F}
F(z, z^*) = \langle {1\over N}\rmtr~\log~(z-\phi)(z^*-\phi^\dgg)\rangle\,,
\eeq
in terms of which\footnote{We use the following notational conventions:
for $z=x+iy$ we define $\pa\equiv {\partial\over \partial z} = {1\over 2}
\left({\partial\over \partial x} -i{\partial\over \partial y}\right)$
so that $\partial z=1$. Similarly, we 
define $\pa^{*}\equiv {\pa\over \pa z^*} = {1\over 2} \left({\pa\over\pa x} + i{\pa\over\pa y}\right)$, so that $\pa^{*}z^*=1$ and also $\pa^*(1/z)=\pi\delta^{(2)}(z)$. Finally, we denote $|z|=r$.}
\beq\label{rho}
\rho(x,y) = {1\over\pi} \partial\pa^{*}~F(z, z^*)\,.
\eeq
The other quantity is the Green's function associated with $\phi$, namely
\beq\label{greens}
G(z,z^*)=\langle {1\over N} \rmtr {1\over z-\phi}\rangle =
\int d^2 x'~{\rho(x', y')\over z-z'}\,,
\eeq
in terms of which 
\beq\label{rho11}
\rho(x,y) = {1\over\pi} \pa^{*}~G(z, z^*)\,.
\eeq

The probability distributions (\ref{prob}) studied in this paper are invariant
under $\phi\rightarrow e^{i\alpha}\phi$, rendering 
\beq\label{radrho}
\rho(x,y)\equiv\rho(r)/2\pi
\eeq
circularly invariant. Rotational invariance thus leads to a simpler form of the defining formula (\ref{greens}) for $G(z, z^*)$ which reads
\beq\label{circular}
\gam(r) \equiv zG(z,z^*)=\int\limits_0^r r'd r'~\rho(r')\,,
\eeq
whence
\beq\label{rhocirc}
\rho(r) = {1\over r} {d\gam\over dr}\,.
\eeq
Clearly, the quantity $\gam(r)$ is a positive monotonically increasing function,
which satisfies the obvious ``sum-rules" 
\beq\label{sumrules}
\gam(0)=0 \quad\quad {\rm and}\quad\quad \gam(\infty)=1\,.
\eeq
In particular, observe that the first condition in (\ref{sumrules}) insures
that no $\delta(x)\delta(y)$ spike arises in $\rho(x,y)$ when calculating it from (\ref{rho11}) with $G(z, z^*)$ given by (\ref{circular}), as it should be. 

We have mentioned in the introduction that a straightforward diagrammatic 
method is not allowed for non-hermitean matrices. The method of 
hermitization\cite{fz} enables us to arrive at a diagrammatic method 
indirectly by reducing the 
problem of determining the eigenvalue density of random non-hermitean matrices to the more elementary problem of determining the eigenvalue density of random hermitean matrices (for which the diagrammatic method may be applied.)
To make the paper self-contained, we summarize the method of hermitization here in the form of an algorithm.

Given a set of random non-hermitean $N\times N$ matrices $\phi$, construct the auxiliary set of random hermitean $2N\times 2N$ matrices 
\beqra\label{H}
H=\left(\begin{array}{cc} 0~~~ & \phi-z\\{} & {}\\
\phi^\dgg-z^* & 0\end{array}\right)\,.
\eeqra
Then one applies one's favorite method of hermitean random matrix theory to 
calculate the propagator associated with $H$, namely, 
\beqra
\cg^{\mu}_{\nu} (\eta; z, z^*) = \langle\left({1\over \eta-H}\right)^{\mu}_{\nu}\rangle\,,
\label{propagator}
\eeqra
where $\eta$ is a complex variable and the indices $\mu$ and $\nu$ run over 
all possible $2N$ values. In particular, in the diagrammatic method, one simply expands $\gmn$ in powers of $1/\eta$, with interaction vertices $H$. This is a well defined procedure for large $\eta$, and it converges to a function which is analytic
in the complex $\eta$ plane, except for the cut (or cuts) along the real axis which contain the eigenvalues of $H$. After summing this series (and thus determining $\gmn (\eta; z, z^*)$ in closed form), we are allowed to set $\eta\rightarrow 0$ in (\ref{propagator}).\footnote{Speaking colloquially, we may say that the crucial maneuver here is that while we cannot expand in powers of $z$, we can arrange for $z$ to ``hitch a ride" with $\eta$ and at the end of
the ride, throw $\eta$ away. }

As mentioned, we can now calculate $\rho(x,y)$ by two different methods, each with its advantages. The first is to simply observe that 
\beqast
-{1\over H} =\left(\begin{array}{cc} 0~~~ & {1\over z^* - \phi^\dgg}\\{} & {}\\
{1\over z- \phi} & 0\end{array}\right)\quad \,, 
\eeqast
from which it follows that 
\beq\label{greens1}
G(z, z^*) = {1\over N} \tr2n \left[\left(\begin{array}{cc} 0~~~ & {\bf 1}_N\\0~~ & 0~\end{array}\right)\cg (0;z, z^*)\right]\,,
\eeq
enabling us to use (\ref{rho11}) to express $\rho(x,y)$. This observation is 
the basis for many of the calculations in \cite{stony}.

An alternative is to take the trace of $\gmn$:
\beq\label{GH}
\cg(\eta; z, z^*) = {1\over 2N} \langle \tr2n~{1\over \eta - H}\rangle
= {\eta\over N}\langle \trn~{1\over \eta^2 - (z^*-\phi^\dgg)(z-\phi)}\rangle\,,
\eeq
from which one can construct an integral representation\cite{fz} of (\ref{F}), thus enabling us to use (\ref{rho}) to determine $\rho(x,y)$. The determination of $\cg(\eta; z, z^*)$ involves only the well-known and
more elementary problem of determining the density of eigenvalues of the
hermitean random matrices $H$, which is the essence of our method of hermitization.

Observe from (\ref{GH}) that 
\beq\label{observe}
\cg(0-i0; z, z^*) = {i\pi\over N} \langle \trn \delta\left(\sqrt{(z^*-\phi^\dgg)(z-\phi)}\right)\rangle
\eeq
and thus counts the (average) number of zero-eigenvalues of the positive semi-definite hermitean matrix $\sqrt{(z^*-\phi^\dgg)(z-\phi)}$ \cite{fz}. 
Thus, if (on the average) $\phi$ has no eigenvalues equal to $z$, or in other words, if the density of eigenvalues of $\phi$ vanishes at $z$, then  
\beq\label{boundary}
\cg(0-i0; z, z^*)= 0\,,
\eeq
independently of the large $N$ limit. In particular, the boundaries of the eigenvalue distribution of $\phi$ in the large $N$ limit, are the curve (or curves) in the complex plane which separate regions where $\cg(0; z, z^*)= 0$
from regions where $\cg(0; z, z^*) \neq 0$. As a matter of fact, one can infer the location of these boundaries even without an explicit knowledge of $\cg(0; z, z^*)$ and $G(z, z^*)$, by investigating the ``gap equation" for $\cg(0; z, z^*)$, namely, the trace of both sides of (\ref{propagator}), and then setting $\eta=0$. The ``gap equation" is a polynomial in $\cg$ with real coefficients (which depend on $z$ explicitly as well as through $G(z, z^*)$, and the parameters of the potential $V$.) Due to the chiral nature of $H$, this polynomial contains a trivial factor
of $\cg(0; z, z^*)$ which we immediately factor out. Setting $\cg =0$ in the remaining factor we obtain an equation for the boundary. This is discussed in detail in \cite{fz}.

To summarize, the method of hermitization allows us to reduce the problem of
dealing with random non-hermitean matrices to the well-studied problem of
dealing with random hermitean matrices. Given the non-hermitean matrix
$\phi$, we study the hermitean matrix $H$ instead. By whatever method one prefers, once one has determined the quark propagator $\gmn$ (or its trace,
the Green's function $\cg (\eta;z,z^{*})$), one can in principle obtain $\rho(x,y)$. Whether that can be done in practice is of course 
another story.

It is worth emphasizing that the formalism described here has nothing to do with large $N$ as such. The formalism is also totally independent of the form of the probability distribution $P(\phi, \phi^\dgg)$. 

\pagebreak

\section{Non-Gaussian Ensembles}
\setcounter{equation}{0}
Our task in this paper is to solve the more problem of determining the 
Green's function and the averaged 
eigenvalue density of a non-hermitean random matrix $\phi$. Here we show that by applying a simple trick,  we can obtain the desired density of eigenvalues with a minimal amount of work, by judiciously exploiting the existing literature on random hermitean matrices.

In this section we will follow a common practice and borrow some terminology from quantum chromodynamics: we may consider $\phi, \phi^{\dgg}$ 
as ``gluons" (in zero space-time dimensions), which interact with a $2N$ 
dimensional multiplet of ``quarks"  $\psi^\mu$, with a complex mass matrix (the ``inverse propagator")
\beqra\label{inverseprop}
\cg_0^{-1} =\left(\begin{array}{cc} \eta~~~ & z\\{} & {}\\
z^* & \eta\end{array}\right)
\eeqra
(expressed in terms of its $N\times N$ blocks.)  
The matrix Green's function $\gmn = \langle\left(1/(\eta-H)\right)^{\mu}_{\nu}\rangle\,,$ in (\ref{propagator}), which represents the quark propagator, is given by the Dyson-Schwinger equation 
in terms of the one-quark irreducible self energy $\smn$:
\beq\label{selfenergy}
\gmn =\left( {1\over {\bf\cg}_0^{-1} - {\bf\Sigma}}\right)^\mu_\nu\,,
\eeq

We emphasize that this equation always holds, and practically amounts to the 
definition of the one-particle irreducible self energy. For probability distributions of the form $P(\phi) = (1/Z) e^{-N\rmtr V(\phi^\dgg\phi)}$ in (\ref{prob}) it is easy to see that\footnote{Split
the $2N$ dimensional quark multiplet into two $N$ dimensional ``flavors", 
$\psi = (u, d)$. Then note that the ``quark-quark-gluon" interaction is 
$\psi^\dgg\left(\begin{array}{cc} 0~~~ & \phi\\\phi^\dgg~ & 0\end{array}\right)\psi = u^\dgg\phi d + d^\dgg\phi^\dgg u $, and use the 
fact that the fermion lines are directed, {\em e.g.} a $\phi$ vertex always
absorbs a $d$ and emits a $u$. Then invoke ``index democracy".}
\beq\label{Sigmamn}
\Sigma^\mu_\nu = \Si(\eta; r) \delta^\mu_\nu
\eeq
(where we recall that $r=|z|$.) Note that $\Si$ depends on $r$ and not separately on $z$ and $z^*$ because it may be expanded as an infinite sum of traces of $\cg_0$, each term of which is a function only of $\eta$ and $r$. Thus,  
\beq\label{gmn}
\gmn = {1\over r^2 - (\eta-\Si)^2} ~
\left(\begin{array}{cc} \Si-\eta ~~~ & z\\{} & {}\\
z^* & \Si-\eta\end{array}\right)\,,
\eeq
leading to 
\beq\label{selfenergy1}
\cg \equiv {1\over 2N}\sum_\mu \cg^\mu_\mu = {\Si - \eta  \over  r^2 - (\eta-\Si)^2}\,.
\eeq

The important point is that in the large $N$ limit, the one-quark irreducible self energy $\Sigma$ can be written in terms of the cumulants $\Gamma_{2k}$ of 
$P(\phi) = (1/Z) e^{-N\rmtr V(\phi^\dgg\phi)}$ (Eq. (\ref{prob})), namely, the connected correlators involving $k$ $\phi$'s and $\phi^\dgg$'s. Diagrammatically speaking, the $\Gamma_{2k}$ may be thought of as a ``blob" out of which emanate $k$ $\phi$'s and $\phi^\dgg$'s, which
cannot be separated into two smaller blobs, with $k_1$ $\phi$'s and $\phi^\dgg$'s in one blob and $k_2$ $\phi$'s and $\phi^\dgg$'s in the other (with $k_1+k_2=k$ of course.) We then obtain  
\beq\label{Sigma}
\Si(\eta; r) = \sum_{k=1}^{\infty} \Gam_{2k} \left[\cg (\eta; r)\right]^{2k-1}\,,
\eeq
in a fashion similar to the calculation in\cite{bluez}. Recall also 
that for the Gaussian ensemble only the variance $\Gamma_2\neq 0$, and thus simply $\Si=\Gamma_2 \cg$.

If we knew $\Si$ in terms of $\cg$, then we can solve for $\cg$ and hence the density of eigenvalues. We are thus faced with the problem of determining 
the $\Gam_{2k}$'s which appears to be a rather difficult task.

It is important to note that the $\Gamma_{2k}$ in (\ref{Sigma}) 
depend only on the probability distribution $P(\phi)$ shown above
and not on the particular quantity we 
average over. Consider the problem of determining the eigenvalue 
density of the hermitean matrix $\pdgp$. We would study the Green's function 
\beq\label{FF}
F(w) = \langle {1\over N} \trn {1\over w -\pdgp}\rangle \equiv \int\limits_0^\infty {\tilde\rho(\si) d\si\over w-\si} \,,
\eeq
where $\tilde\rho(\mu)$ is the averaged eigenvalue density of $\pdgp$. 
We see from (\ref{GH}) that this Green's function is related to $\cg (\eta ; r=0)$ simply through
\beq\label{relation} 
\cg (\eta ; r=0) = \eta F( \eta^2 )\,.
\eeq
Also, for $z=0$ we have from (\ref{selfenergy1})
\beq\label{selfenergy2}
\Si (\eta; 0) = \eta - \cg^{-1}(\eta; 0)\,.
\eeq

The crucial observation is that $F(w)$ is already known in the literature on chiral and rectangular block random hermitean matrices\footnote{Namely, matrices $H$ with the block structure 
$H=\left(\begin{array}{cc} 0~~~ & \phi\\
\phi^\dgg & 0\end{array}\right)\,.$}
for the Gaussian distribution\cite{bhz, rectangles, bhznpb, rectangles1}, as well as 
for non-Gaussian probability distributions of the form (\ref{prob}) with an arbitrary polynomial potential $V(\pdgp)$)\cite{ambjorn, periwal}.
In fact, the authors of \cite{ambjorn, periwal} simply calculated the diagonal 
elements of the propagator $\gmn (\eta;0)$ using Dyson gas techniques (so the 
coefficients $\Gam_{2k}$ are only implicit in these papers.)

The contents of (\ref{Sigma}), (\ref{relation}) and (\ref{selfenergy2}) may 
thus be summarized as follows: there is a unique function $ a (\xi)$ which behaves like $1/\xi$ as $\xi$ tends to infinity, is regular at $\xi=0$, and satisfies the equation $\xi-a^{-1}=\sum_{k=1}^{\infty} \Gam_{2k} a^{2k-1}$. 
This function is 
\beq\label{a}
a(\xi) = \xi F(\xi^2) \equiv \cg(\xi; 0)\,. 
\eeq

We, on the other hand, are interested in the opposite case, where $\eta=0, r=|z|\neq 0$. Consider the matrix $\gmn (0-i0; z, z^*)$. Due to `` color index democracy",
its $N\times N$ blocks are all proportional to the unit matrix ${\bf 1}_N$, and we see that 
\beq\label{prop0}
\gmn (0-i0; z, z^*) =\left(\begin{array}{cc} \cg~ & G^*\\{} & {}\\
G~ & \cg~\end{array}\right)\,,
\eeq
where $\cg(0-i0; r)$ is given in (\ref{observe}) and the complex Green's function $G(z, z^*)$ was defined \footnote{From this point on, unless otherwise stated, our notation will be such that $\cg\equiv\cg(0-i0;r)$, $\Si\equiv\Si(0;r)$ and $G\equiv G(z, z^*)$. Note in particular from (\ref{observe}) that $\cg$ is then pure imaginary, 
with $\Im \cg\geq 0$.} in (\ref{greens1}). Inverting the propagator in (\ref{prop0}) and using (\ref{inverseprop}) (with $\eta=0$) and (\ref{selfenergy}), we obtain the two equations
\beq\label{zsi1}
z = {G^*\over |G|^2 - \cg^2}
\eeq
and 
\beq\label{zsi11}
\Si = {\cg\over |G|^2 - \cg^2}\,.
\eeq

Note from (\ref{zsi1}) that if $\cg=0$, then $G=1/z$, which corresponds to the region outside the domain of the eigenvalue distribution\cite{stony,fz}.
In addition, using (\ref{zsi1}) twice, we obtain
\beq\label{zg}
zG = {|G|^2\over |G|^2-\cg^2} = (|G|^2-\cg^2)~r^2\,.
\eeq
Thus, due to the fact that $\cg$ is pure imaginary (see (\ref{observe}))
we conclude that $\gam\equiv zG$ is always real and non-negative.
Alternatively, we can write 
\beq\label{gg}
\cg^2 = {zG~(zG-1)\over r^2}\,,
\eeq
and therefore from the fact that $\cg^2\leq 0$ we conclude that $0\leq\gam\leq 1$,
in accordance with the monotonicity of $\gam (r)$ and the sum-rules $\gam(0)=0$ and $\gam(\infty)=1$ in (\ref{sumrules}).

Let us now define the pure imaginary quantity
\beq\label{xi1}
\xi\equiv {1\over \cg} + \Si = {1\over \cg} + \sum_{k=1}^{\infty} \Gam_{2k} \cg^{2k-1}\,,
\eeq
where the last equality follows from (\ref{Sigma}) (with $\eta=0$).
Using (\ref{zsi1}) and (\ref{zsi11}) we thus have simply 
\beq\label{final11}
\xi= {zG\over \cg}\,.
\eeq
It then follows from (\ref{a}) and the statement preceding that equation, and from (\ref{final11}) that 
\beq\label{final12}
\xi F(\xi^2) = \cg = {zG\over\xi}\,.
\eeq
We have thus eliminated the $\Gam_{2k}$ !
 
From (\ref{gg}) and from (\ref{final11}) we determine $\xi=\gam/\cg$ to be
\beq\label{xi}
\xi^2 = {\gam \over \gam -1}~r^2\,.
\eeq
Note that $\xi^2\leq 0$, consistent with the bounds $0\leq\gam\leq 1$.
Comparing (\ref{final12}) and (\ref{relation}) we obtain the remarkable 
relation 
\beq\label{remarkablerelation} 
\cg(\xi;0) = \cg(0;r)\,,
\eeq
for $\xi$ and $r$ that are related by (\ref{xi}) (with $\xi$ imaginary.)

We now have the desired equation for $\gam\equiv zG$: substituting (\ref{xi}) into (\ref{final12}) we obtain
\beq\label{final}
\gam\left[r^2~F\left({\gam~r^2\over \gam -1}\right) -\gam +1\right] = 0\,.
\eeq
Thus, given $F$ we can solve for $\gam(r)$ using this equation.

From (\ref{FF}) we observe that for $w$ large we may expand
\beq\label{expand}
F(w) = {1\over w} + {1\over w^2} f(w)\,,
\eeq
where $f(w) = f_0 + (f_1/w) + \cdots $. Substituting 
$w=\xi^2=\gam r^2/(\gam-1)$ into (\ref{expand}) and letting $\gam$ tend to $1$, (\ref{final}) becomes
\beq\label{doubleroot}
(\gam-1)^2\left[{f(w)\over \gam r^2} -1\right]\,,
\eeq
and thus $\gam=1$, which corresponds to the region outside the eigenvalue distribution, is always a double root of (\ref{final}).

We will now explain how the Green's function $F(w)$ is obtained in the literature\cite{rectangles, ambjorn, periwal}. Let $V$ be a polynomial of 
degree $p$. From the definition of $F(w)$ in (\ref{FF}) and its 
analyticity property, we expect $F$, following the arguments of Br\'ezin et al,
to have the form\footnote{Here we assume for simplicity that the eigenvalues of $\pdgp$ condense into a single segment $[a,b]$.}
\beq\label{FFF}
F(w)={1\over 2}V'(w)-P(w)\sqrt{(w-a)(w-b)}\,,
\eeq
where 
\beq\label{P}
P(w) = \sum_{k=-1}^{p-2} a_k~w^k\,.
\eeq
The constants $0 \leq a < b$ and $a_k$ are then determined completely
by the requirement that $F(w)\rightarrow {1\over w}$ as $w$ tends to infinity, and by the condition that $F(w)$ has at most an integrable 
singularity as $w\rightarrow 0$.~\footnote{Thus, if $a>0$, inevitably $a_{-1} = 0$. However, if $a=0$, then $a_{-1}$ will be determined by the first condition.}

\pagebreak

\section{Phase Transitions in the Quartic Ensemble}
\setcounter{equation}{0}

Having determined $F(w)$ in this way, we substitute it into (\ref{final}) and find $G(z, z^*)$. We can thus calculate the density of eigenvalues explicitly for an arbitrary $V$. For the sake of simplicity we now focus on the quartic potential\footnote{In the Gaussian case, $V=\pdgp$, we have $2\xi F(\xi^2) = \xi - \sqrt{\xi^2 -4}$, whence the roots of (\ref{final}) are $\gam=0, 1$ and $r^2$. We note that $\gam=0$ is unphysical, $\gam=r^2$
({\em i.e.,} $G=z^*$) corresponds to Ginibre's disk\cite{ginibre}, and $\gam=1$
is the solution outside the disk.}
\beq\label{quartic}
V(\pdgp) = 2 m^2\pdgp + g (\pdgp)^2\,.
\eeq
\subsection{The Disk Phase}
For $m^2$ not too negative, which as we shall see translates here into
$m^2>-\sqrt{2g}$, we expect the density of eigenvalues to be a disk.
In that case we expect that $\rho(0)>0$, as in Ginibre's case. Thus, from (\ref{rhocirc}) $\rho(r) = (1/r) (d\gam/dr)$ and from the first sum rule $\gam(0)=0$ in (\ref{sumrules}) we conclude that \footnote{Alternatively, we 
can reach the same conclusion about $\gam (r)$ for $r$ small by looking at $\cg^2$. According to (\ref{boundary}), $\cg^2$ must not vanish in some neighborhood of the origin. Neither can it diverge there. It thus follows from  (\ref{gg}) that $\gam(r)\sim  r^2$ near $r=0$.} $\gam(r)\sim  r^2$ near $r=0$. Therefore, for $r$ small, (\ref{final}) yields $F(-r^4)\sim -1/r^2$, namely, $F(w)\sim 1/\sqrt{w}$
for $w\sim 0$, as we could have anticipated from Ginibre's case. This means 
that $a=0$ in (\ref{FFF}), and therefore, 
\beq\label{FQ}
F(w)=m^2 + gw - \left({a_{-1}\over w} + a_0\right)~\sqrt{w (w-b)}\,.
\eeq
From the asymptotic behavior of $F(w)$ we then obtain\footnote{$b$ is positive by definition. The expression for $b$ in (\ref{parameters1}) is the positive root (also for $m^2<0$) of the quadratic equation which determines $b$. $b$ then determines $c$ via $c=m^2 + (gb/2)$.}
\beq\label{parameters1}
a_0=g, \quad c\equiv a_{-1}={2m^2 +\sqrt{m^4 + 6g}\over 3}, \quad {\rm  and}\quad b={-2m^2 + 2\sqrt{m^4 + 6g}\over 3g}\,.
\eeq
For later convenience we record here that 
\beq\label{bc2}
b c^2 = 2{(m^4+6g)^{3/2}-m^2(m^4-18g)\over 27 g}\,.
\eeq

Substituting (\ref{FQ}) and (\ref{parameters1}) into (\ref{final}) we arrive\footnote{In deriving (\ref{disk}) we divided through by the trivial factor $(\gam-1)^2$ alluded to in (\ref{doubleroot}).} at
\beq\label{disk}
\gam^2 - \gam~[1+2(m^2r^2+gr^4)] + bc^2r^2 =0
\eeq
with roots
\beq\label{gdisk+-}
\gam_\pm(r) = {1\over 2}\left[1+2(m^2r^2+gr^4) \pm 
\sqrt{[1+2(m^2r^2+gr^4)]^2 - 4bc^2r^2}\,\right]\,.
\eeq

Let us assume that $m^2>-\sqrt{2g}$. Under this condition the the discriminant of (\ref{disk}) is non-negative, and contains a factor $(r^2-b/4)^2$. The root of (\ref{disk}) which satisfies the condition $\gam(0)=0$ (see (\ref{sumrules})) is $\gam_-(r)$ (whereas $\gam_+(0)=1$.)  However, a subtlety arises, since $\gam_-(r)$ is monotonically increasing (as 
it should be, by definition) only for $r^2<b/4$. The root $\gam_+(r)$ increases monotonically only for $r^2>b/4$. Due to the factor $(r^2-b/4)^2$ in the
discriminant of (\ref{disk}), we observe that $\gam_-^{'}(\sqrt{b}/2 -)=\gam_+^{'}(\sqrt{b}/2 +)$, and thus $\gam_+$ and $\gam_-$ may be glued smoothly into a single monotonically increasing function $\gam(r)_{disk}$ as
\beq\label{gdisk}
\gam(r)_{disk} = \left\{\begin{array}{c} \gam_-(r)\quad ,\quad r^2<b/4\,\,\\
\gam_+(r)\quad ,\quad r^2>b/4\,.\end{array}\right.
\eeq

The eigenvalue density is then given by (\ref{rhocirc}) as
\beq\label{rhodisk}
\pi\rho(x,y) = m^2 + 2gr^2 + [{\rm sgn}~({b\over 4}-r^2)] ~{bc^2-(m^2+2gr^2)[1+2(m^2r^2+gr^4)]\over\sqrt{[1+2(m^2r^2+gr^4)]^2 - 4bc^2r^2}}\,.
\eeq
In particular, note that $\rho(0,0)=bc^2/\pi\geq 0$.

We now determine the radius of the disk $r_b$, namely, the boundary of the eigenvalue distribution. In this case there is only one boundary, and thus the $r$ dependent solution of (\ref{final}) increases monotonically from $\gam(0)=0$ to $\gam(r_b)=1$, since all the eigenvalues are included within $r\leq r_b$.
Thus, setting $\gam(r)=1$ in (\ref{disk}) and using (\ref{bc2}) we obtain 
\beq\label{rb}
r_b^2 = (bc^2 - 2m^2)/2g = 2{(m^4+6g)^{3/2}-m^2(m^4 + 9g)\over 27 g}\,.
\eeq
It is possible to show that $r_b^2> b/4$. We thus find that the density at the boundary is 
\beq\label{rhob}
\pi\rho_{b} = \left[ 2r_b^2 +{2m^2\over g} - {1\over 2 g r_b^2} \right]^{-1}
\eeq
which can be shown to be always positive.

In the limit $g\rightarrow 0$ (with $m^2>0$, of course) $c$ tends to $m^2$ and both $r_b^2$ and $b/4$  tend to $1/(2m^2)$. Therefore, (\ref{gdisk}) simplifies into
\beqast
\gam(r)_{disk} = \left\{\begin{array}{c} 2 m^2 r^2 \quad ,\quad r^2<1/(2m^2)\,\,\\
1\quad ,\quad r^2>1/(2m^2)\,.\end{array}\right.
\eeqast
Thus, from (\ref{rhocirc}) (or by taking $g\rightarrow 0$ in (\ref{rhodisk}))
we recover Ginibre's uniform density $\rho(x,y) = 2m^2/\pi$ inside a disk of radius $1/(2m^2)$, consistent with the $g\rightarrow 0$ limit of (\ref{quartic}).

In the strong coupling limit $g/m^4 >>1$ we observe that $c\sim \sqrt{2g/3}$ and $b\sim\sqrt{8/3g}$.
Therefore, in this limit the disk shrinks like $r_b^2\sim \sqrt{8/27 g}$, which means that the density should blow up like $1/r_b^2$. We find indeed that in 
this limit  $\rho_{b}\sim (12/5\pi)\sqrt{6g}$. We also find that in this limit $\rho(0,0)\sim (2/\pi)(2/3)^{3/2}\sqrt{g}\sim (5/27) \rho_{b} < \rho_b$, and the density grows as $r$ increases. 

To summarize, for $m^2>-\sqrt{2g}$, the $\gam(r)$ corresponding to the 
ensemble (\ref{quartic}) in the disk phase is thus given by (\ref{gdisk}) for $r\leq r_b$ and is equal to one for $r>r_b$.

When $m^2$ becomes too negative, that is, for $m^2 <-\sqrt{2g}$, the discriminant of (\ref{disk}) becomes negative for some range of values of $r$, rendering $\gam_\pm(r)$ complex, which is unacceptable. This signals that the disk phase becomes unstable, and the systems undergoes a phase transition, as we discuss in the next section. The boundary of the disk phase in the $m^2-g$ 
plane is thus given by $m^2 = -\sqrt{2g}$. Consider approaching this boundary from within the disk phase by setting $m^2=-\sqrt{2g} + \delta$, with $\delta$
positive and small. Then, using (\ref{parameters1}), we find to first order in $\delta$ that $c=\delta/2$ and $b=2\sqrt{(2/g)} - \delta/g$.
In particular, at the phase boundary itself $c=0$. Consequently, at the phase boundary (\ref{rhodisk}) simplifies into 
\beq\label{rhodiskcrit}
\rho (x,y) = \left\{\begin{array}{c} 0\quad ,\quad r^2<1/\sqrt{2g}\,\,\\
2\sqrt{2g} (\sqrt{2g} r^2 -1) \quad ,\quad 1/\sqrt{2g}<r^2<r_b^2\,.\end{array}\right.
\eeq
where $r_b^2=\sqrt{2/g}$ from (\ref{rb}). Thus, as we decrease $\delta$ to zero, $\rho(x,y)$ becomes increasingly depleted inside the disk $r^2< b(\delta)/4$, 
reaching complete depletion at $\delta=0$, at which point the disk breaks into
an annulus. Finally, for later reference, let us note that at the phase boundary (\ref{FQ}) reads
\beq\label{FQcrit}
F(w)=-\sqrt{2g} + gw - g~\sqrt{w \left(w-2\sqrt{{2\over g}}\right)}\,.
\eeq

\subsection{The Annular Phase}

We saw that at the phase boundary $m^2=-\sqrt{2g}$ the disk configuration of 
the previous sub-section broke into an annulus. We thus expect that for 
$m^2 <-\sqrt{2g}$ the stable eigenvalue distribution would be annular. 

For convenience, let us switch notations according to $m^2=-\mu^2$, and also 
write $\mu_c^2=\sqrt{2g}$.

We denote the radii of the annulus boundaries by $r_1<r_2$.
In the inner void, $r<r_1$, we clearly have $\gam(r)\equiv 0$. Thus, (\ref{final}) and (\ref{FFF}) imply $a_{-1}\sqrt{ab}=0$. The annulus must therefore arise for $a_{-1} = 0$ (the other possible solution $a=0, a_{-1}\neq 0$ leads to a disk configuration with $\gam=0$ only at $r=0$, as we discussed in the previous sub-section.) Therefore, 
\beq\label{FANN}
F(w)=m^2 + gw - a_0~\sqrt{(w-a) (w-b)}\,.
\eeq
From the asymptotic behavior of $F(w)$ we then obtain 
\beq\label{parameters2}
a_0=g, \quad a={\mu^2\over g} -\sqrt{{2\over g}}, \quad {\rm  and}\quad b={\mu^2\over g} +\sqrt{{2\over g}}\,.
\eeq
We see that $a= (2/\mu_c^4)(\mu^2-\mu_c^2)$ which is positive for $\mu^2>\mu_c^2$, as it should be, by definition.

We now determine $\gam(r)$ inside the annulus. Substituting (\ref{FANN}) and (\ref{parameters2}) into (\ref{final}) and factoring out
all the factors \footnote{There is one extra such factor as compared to the 
disk phase.} of $(\gam-1)$, we obtain a linear equation for $\gam$ and find
\beq\label{gann}
\gam(r)_{annulus} = 1-2\mu^2r^2+2gr^4\,,
\eeq
which must correspond to the region inside the annulus.

The $\gam(r)$ configuration (\ref{gann}) has to match continuously to 
$\gam=0$ at $r=r_1$ and to $\gam=1$ at $r=r_2$. From the matching at 
$r=r_2$ we find that
\beq\label{r2}
r_2^2={\mu^2\over g} = {2\mu^2\over \mu_c^4}\,.
\eeq
The matching at $r=r_1$ yields a quadratic equation for $r_1^2$ with roots
\beq\label{tworoots}
{\mu^2\pm\sqrt{\mu^4-2g}\over 2g} = {\mu^2\pm\sqrt{\mu^4-\mu_c^4}\over \mu^4_c}\,,
\eeq
both of which are smaller than the expression in (\ref{r2}). In order to determine which of these roots corresponds to $r_1^2$ we recall that 
$\gam(r)$ must interpolate monotonically between its two boundary values, 
namely, that $\gam'(r)>0$ for $r_1<r<r_2$, which is nothing but the statement that $\rho(r)>0$. Substituting (\ref{gann}) in (\ref{rhocirc}) we find 
\beq\label{rhoann}
\rho(x,y) = {4g\over\pi}\left(r^2 -{\mu^2\over\mu_c^4}\right) = {2\mu_c^2\over\pi}\left(\mu_c^2r^2 -{\mu^2\over\mu_c^2}\right)\,,\quad\quad r_1<r<r_2\,.
\eeq
Note that $\rho(r)$ increases monotonically. The correct
normalization of (\ref{rhoann}) is guaranteed by $\gam(r_2)=1$, as can
be checked explicitly.
Finally, positivity of (\ref{rhoann}) determines $r_1^2$ in (\ref{tworoots})
as
\beq\label{r1}
r_1^2={\mu^2 + \sqrt{\mu^4-\mu_c^4}\over \mu^4_c}\,.
\eeq
Note that $r_1^2=1/\mu_c^2=1/\sqrt{2g}$ at $\mu^2=\mu^2_c$, which coincides 
with the radius (squared) of the depleted region in the disk configuration (\ref{rhodiskcrit}). Also, at $\mu^2=\mu^2_c$ we have $r_2^2 = 2/\mu_c^2 = \sqrt{2/g}$, which coincides with the disk radius squared $r_b^2$ in (\ref{rb}) at the phase boundary. Thus, at the phase boundary $\mu^2=\mu_c^2$ 
(\ref{rhoann}) coincides with (\ref{rhodiskcrit}), namely, $\rho(x,y)$
is continuous at the transition from the disk phase to the annular phase.

Consider approaching the phase boundary $m^2 = -\sqrt{2g}$ from within the 
annular phase by setting $\mu^2=\sqrt{2g} + \delta$, with $\delta$
positive and small. Then, since all the expressions in (\ref{parameters2}) are
linear in $\mu^2$, we find that $a=\delta/g$ and $b=2\sqrt{(2/g)} + \delta/g$.
Thus, at the phase boundary itself $a=0$, and $b=1/\sqrt{2g}$. Therefore, 
at the phase boundary (\ref{FANN}) reads
\beq\label{ANNcrit}
F(w)=-\sqrt{2g} + gw - g~\sqrt{w \left(w-2\sqrt{{2\over g}}\right)}\,,
\eeq
which coincides with (\ref{FQcrit}). Thus, $F(w)$ (and consequently, the eigenvalue density of $\phi^\dgg\phi$) is also continuous at the transition.

It is amusing to note that the various expressions we encountered in this sub-section are much simpler than their counterparts in the disk phase.

\pagebreak

\section{Boundaries and Shape Universality of the Eigenvalue Distribution:
the Single Ring Theorem}
\setcounter{equation}{0}

In the previous section we analyzed the probability distribution (\ref{prob})
with a quartic potential $V(\pdgp) = m^2\pdgp + {g\over 2}(\pdgp)^2$.
We obtained circularly symmetric eigenvalue distributions: a disk (\ref{rhodisk}) and an annulus (\ref{rhoann}). Making $m^2$ negative enough drove a phase transition from the disk configuration to the annulus configuration.

This state of affairs obviously generalizes. In this class of models, 
the domain of the eigenvalue distribution is always circularly symmetric, due 
to the rotational invariance of the probability distribution (\ref{prob}).
Consider a potential $V(\pdgp)$ with several wells or minima. For deep enough
wells, we expect the eigenvalues of $\pdgp$ to ``fall into the wells".
Thus, one might suppose that the eigenvalue distribution of $\phi$ to be 
bounded by a set of concentric circles of radii $0\leq r_1 < r_2 < \cdots < r_{n_{\rm max}}$, separating annular regions on which $\rho(r)>0$ from voids (annuli in which $\rho(r)=0$.) A priori, it is natural to assume that the maximal number of such circular boundaries should grow with the degree of $V$, because $V$ may then have many deep minima.

Remarkably, however, we prove below 
that the number of these boundaries is two at the most. To reconcile this conclusion with the a priori expectation just mentioned, we note that 
while the eigenvalues of the hermitean matrix $\pdgp$ may split into several disjoint segments along the positive real axis, this does not necessarily constrains the eigenvalues of $\phi$ itself to condense into annuli. This statement is made more precise in the Appendix.

We call this result the ``Single Ring Theorem." To prove it, let us assume for the moment that there are $n$  boundaries.  It is easy to see from (\ref{circular}), the defining equation of $\gam(r)$, that in the annular void $r_k<r<r_{k+1}$ in the eigenvalue distribution, $\gam$ is a constant which is equal to the fraction of eigenvalues contained inside the 
disk $r\leq r_k$. Thus, the equation to determine $\gam$ (Eq. (\ref{final})
which we repeat here for convenience)
\beq\label{final6}
\gam\left[r^2~F\left({\gam~r^2\over \gam -1}\right) -\gam +1\right] = 0
\eeq
must have a series of monotonically increasing constant solutions $\gam_1<\gam_2<\cdots \leq 1$, which correspond to the various voids.

In particular, from (\ref{sumrules}) we have $\gam =1$ for $r>r_{n_{\rm max}}$, namely, $G=1/z$. Thus, $\gam=1$ must be a solution of (\ref{final6}), the maximal allowed constant solution. It is straightforward 
to see that $\gam=1$ is indeed a solution, because we know that 
\beqast
F\left({\gam \over \gam -1}~r^2\right) \sim {\gam -1\over \gam r^2}
\eeqast
as $\gam$ tends to $1$. Also, if $r_1>0$, namely,
if there is a hole at the center of the eigenvalue distribution, then for
$r\leq r_1$ we must have $\gam =0$ (independently of $r$), which is obviously a solution\footnote{From the paragraph right below (\ref{FFF}) it is clear that $w F(w)\rightarrow 0$ at $w=0$.} of (\ref{final6}). (On the other hand, if $r_1=0$, so that the eigenvalue distribution includes the origin, an $r$ independent solution $\gam=0$ is of course only a spurious solution which should be discarded.)

Assume now that $\gam=\gam_0$ is an $r$ independent solution of (\ref{final6}).
Taking the derivative of (\ref{final6}) with respect to $r^2$ at $\gam\equiv\gam_0$ we obtain 
\beq\label{gam0}
{\gam_0\over\gam_0-1}\left[ F(\xi^2) + \xi^2{dF\over d\xi^2}\right]_{|_{\xi^2={\gam_0 \over \gam_0 -1}r^2}} =0\,,
\eeq
which is the condition for the existence of an $r$ independent solution $\gam_0$. Thus, there are two possibilities: either $\gam_0=0$, which we 
already encountered, or $ F(\xi^2) + \xi^2{dF\over d\xi^2}=0$. This equation
immediately yields
\beq\label{Fxi}
F(\xi^2)={1\over \xi^2}\,,
\eeq
where the integration constant is fixed by the asymptotic behavior of $F(\xi^2)$
as $\xi\rightarrow\infty$. But for a generic $V(\pdgp)$, $F(w)$ is given by
(Eq. (\ref{FF}))
\beqast
F(w) = \langle {1\over N} \trn {1\over w -\pdgp}\rangle \equiv \int\limits_0^\infty {\tilde\rho(\si) d\si\over w-\si} \,,
\eeqast
with (\ref{Fxi}) being the asymptotic behavior of $F(w)$ as
$w=\xi^2\rightarrow\infty$. We thus conclude that $\xi^2\rightarrow\infty$
in (\ref{Fxi}), namely, that $\gam_0=1$. Thus, to summarize, the only 
possible $r$ independent solutions of (\ref{final}) are $\gam=0$ and $\gam=1$, which we already discussed. Since no other $r$ independent solutions arise,
there can be no more than a single void in the eigenvalue distribution,
whatever polynomial the potential $V(\pdgp)$ is. The shape of the eigenvalue distribution is thus either a disk or an annulus.

Interestingly enough, we can arrive at the same conclusion by invoking other general aspects of the method of hermitization \cite{fz}, and thus providing a nice self consistency check of our formalism. Recall at this point that the boundaries of the eigenvalue distribution are given by (\ref{boundary}), namely, the zeros of $\cg$. But $\cg$ is given by (Eq. (\ref{gg}))
\beqast
\cg^2 = {\gam (\gam-1)\over r^2}\,,
\eeqast
from which we see that at the boundaries $\gam$ can take on only two values : zero and one. Since $\gam$ is a constant in the void, by continuity, these are the only possible values 
of $\gam$ inside any of the voids. Therefore, there may be two circular
boundaries at most. In other words, as far as the eigenvalue density of $\phi$ is concerned, an ensemble of the form (\ref{prob})
may have two phases at most, as we concluded earlier.

In addition to $\gam = 1$ (and possibly $\gam = 0$), (\ref{final6}) 
must have other roots which do depend on $r$. Among all these other roots, we expect to find a unique root $\gam(r)$, which is a positive monotonically increasing function of $r$,  that matches continuously at the boundaries $r_k\quad (k\leq 2)$ to the constant roots of (\ref{final6}). The actual $zG(z, z^*)$ of the ensemble (\ref{prob}) is therefore a continuous monotonically increasing function of $r$, as required by (\ref{circular}).
If the eigenvalue distribution is annular with boundaries $0\leq r_1<r_2$ (the case $r_1 =0$ corresponding to the full disk), it vanishes for $0\leq r\leq r_1$, rises monotonically from zero to one on $r_1\leq r\leq r_2$, and equals 
to one for $r\geq r_2$.

\pagebreak
 
\section{The Addition Formalism for non-Hermitean Random Matrices}
\setcounter{equation}{0}
Using the method of hermitization\cite{fz}, we develop in this section
the addition formalism for non-hermitean random matrices. Our results agree
with recent results of Zahed {\em et al.}\cite{stony}. We feel that 
the method of hermitization clearly elucidates the development of this 
addition formalism. At the end of this section we use the addition formalism 
to formulate and prove a non-hermitean non-Gaussian version of the central 
limit theorem.

Let $\phi_1$ and $\phi_2$ be two $N\times N$ random non-hermitean
matrices, drawn independently from the probability distributions $P_1(\phi_1^\dgg\phi_1)$ and $P_2(\phi_2^\dgg\phi_2)$. Furthermore, let $H_a\,\,(a=1,2)$ be the $2N\times 2N$ hermitean Hamiltonians associated with $\phi_a$, as defined by (\ref{H}),
and similarly define the ``total Hamiltonian" as
\beq\label{H12}
H_{T} = \left(\begin{array}{cc} 0~~~ & \phi_1+\phi_2-z\\{} & {}\\
\phi_1^\dgg+\phi_2^\dgg-z^* & 0\end{array}\right)\,.
\eeq
The addition problem for $\phi_1$ and $\phi_2$ may be formulated as follows: Consider the probability distributions
\beq\label{propa}
P_a(\phi_a^\dgg\phi_a) = {1\over Z_a}~{\rm exp}~\left[-N\rmtr V_a(\phi_a^\dgg\phi_a)\right]\,,\quad\quad a=1,2\,.
\eeq
Given the two propagators 
\beqra
\gamn (\eta; z, z^*) = \langle\left({1\over \eta-H_a}\right)^{\mu}_{\nu}\rangle^{(a)}
\label{propagatora}
\eeqra
(where $\langle \cdots\rangle^{(a)}$ indicates an average with respect to 
$P_a$), find (in the large $N$ limit) the propagator
\beqra
\cg^{\mu}_{\nu} (\eta; z, z^*) = \langle\left({1\over \eta-H_{T}}\right)^{\mu}_{\nu}\rangle
\label{prop12}
\eeqra
associated with the probability distribution 
\beq\label{prob12}
P(\phi_1^\dgg\phi_1, \phi_2^\dgg\phi_2) = P_1(\phi_1^\dgg\phi_1)\, P_2(\phi_2^\dgg\phi_2)\,.
\eeq 

The propagators $\gamn (\eta; z, z^*)$ may be obtained in principle, by 
applying the formalism of Section 3. Once one has succeeded in solving the addition problem and obtained an explicit expression for (\ref{prop12}), one 
may set $\eta=0$ and calculate the various averages $F(z, z^*) = \langle {1\over N}\rmtr~\log (z-\phi)(z^*-\phi^\dgg)\rangle\,, G(z, z^*) = \langle (1/N) ~\rmtr ~1/(z-\phi)\rangle$ and $\cg(0;z, z^*)$ as prescribed 
in Section 2, from which one may further calculate the averaged density of eigenvalues $\rho(x, y)$ of $\phi_1+\phi_2$.

The formulation of the addition problem for non-hermitean random matrices $\phi_1, \phi_2$  bears obvious similarities to the formulation of 
the addition problem for hermitean random matrices. In fact, if in the formulation just given one replaces the bare inverse propagator $\cg_0^{-1}$
defined in (\ref{inverseprop}) simply by $\cg_0^{-1}\equiv z$, and the chiral matrices 
\beqast\left(\begin{array}{cc} 0~~~ & \phi_a\\{} & {}\\
\phi^\dgg_a & 0\end{array}\right)\eeqast
by the ordinary hermitean matrices $\varphi_a \quad(a=1,2)$, one 
would recover the formulation of the usual addition problem for hermitean 
random matrices\cite{blue, bluez, zahed}. These identifications of various
quantities that appear in the formulations of the two addition problems, enable us to derive the addition formalism for random non-hermitean matrices by
following the steps of the by now well known addition formalism for ordinary
hermitean random matrices\cite{bluez}, which we now proceed to do.

Let $\Gak$ and
\beq\label{Sigmamna}
\siamn = \sia(\eta; r) \delta^\mu_\nu
\eeq
be, respectively, the connected $2k$ gluon ``blob" (or cumulant)
and the self-energy associated with the probability distribution $P_a(\phi_a^\dgg\phi_a)$. From the definition of self-energy
\beqast
\gmn =\left( {1\over {\bf\cg}_0^{-1} - {\bf\Sigma}}\right)^\mu_\nu
\eeqast
(Eq. (\ref{selfenergy})) we find 
\beq\label{selfenergya}
\siamn = \left(\invp - \cg^{-1(a)}\right)^\mu_\nu\,. 
\eeq
In addition, rewriting the expansion (\ref{Sigma}) for $\siamn$ we have 
\beq\label{Sigmaa}
\sia (\eta; r) = \sum_{k=1}^{\infty} \Gak \left[{1\over 2N}~\gamm (\eta; z,z^*)\right]^{2k-1} \equiv \sum_{k=1}^{\infty} \Gak \left[\ga (\eta; r)\right]^{2k-1}\,.
\eeq
For reasons that will become clear below we bothered to display the trace operation in (\ref{Sigmaa}) explicitly.
The two gluon types $\phi_1$ and $\phi_2$ in the combined probability distribution (\ref{prob12}) do not interact and therfore the self-energy 
$\smn = \Si (\eta; r) \delta_\nu^\mu$ associated with (\ref{prob12}) 
is given by 
\beq\label{Sigmatot}
\Sigma (\eta; r) = \sum_{k=1}^{\infty} \left(\sum_a\Gak\right) \left[{1\over 2N}~\gmm (\eta; z,z^*)\right]^{2k-1}\equiv\sum_{k=1}^{\infty} \left(\sum_a\Gak\right)\left[ \cg (\eta; r)\right]^{2k-1}\,,
\eeq
where now obviously the internal quark lines are given by (\ref{prop12}).

The right hand side of (\ref{Sigmaa}) and of (\ref{Sigmatot}) are expansions
in powers of the trace of three different propagators: $\gamn~(a=1,2)$ and
$\gmn$. These three propagators are obviously not the same, but we observe that if they were identical, then (\ref{Sigmatot}) would simply be the sum (over the two values of $a$) of (\ref{Sigmaa}). The fact that these three propagators are not identical renders the relation between the three self-energies more complicated. The essence of the addition formalism is to maneuver around this difficulty. The idea is as follows: for the sake of argument, assume 
that it is mathematically consistent to consider $\gamn (\eta;, r)$ as a matrix valued function of the inverse propagator matrix
\beqast
\cg_0^{-1} =\left(\begin{array}{cc} \eta~~~ & z\\{} & {}\\
z^* & \eta\end{array}\right)
\eeqast
(Eq. (\ref{inverseprop}).) This function would then have a matrix valued functional inverse $\cb_a$, where $\cb_a\left(\cg\left(\cg_0^{-1}\right)\right)=\cg_0^{-1}$. Naively, we may deduce from (\ref{selfenergya})
that
\beq\label{naiveblue}
\cb_a(\cg^{(a)\alpha}_\beta)^\mu_\nu = \left(\cg^{-1(a)}\right)^\mu_\nu + \siamn
\eeq
(below we shall do away with the adverb ``naively" and make the construction of $\cb_a$ mathematically precise.) Using this matrix valued 
functional inverse we would be able to solve the equation $\gamn (\cx_a) = 
\gmn$ (where $\cx_a$ is the matricial argumet of the matrix valued function $\gamn$  ) as $\cx_a = \cb_a(\gmn)$. We would then evaluate each of the $\sia$ in
(\ref{Sigmaa}) at the appropriate argument $\cx_a$ (instead at argument $\invp$). The right hand side of (\ref{Sigmaa}) would then be an expansion in powers of $\cg (\eta; r)$, and we would thus conclude that $\sum_a \sia (\cx_a) = \Si (\invp)$. 

The arguments in the last paragraph follow the same logic as in the case of 
the addition formalism of random hermitean matrices. Recall that in 
the theory of ordinary hermitean random matrices, the Blues function  
$\cb(w)$ and the Green's function $\cg(z)\equiv (1/N)\langle \rmtr~(z-\varphi)^{-1}\rangle$ are functional inverses of one another\cite{bluez}. In the present context, we already noted that the ordinary
complex variable $z$ is replaced by the matrix $\invp$ in (\ref{inverseprop}), which leads us to consider matrix valued functions of $\invp$ and composition thereof. Thus, following the nomenclature of the hermitean addition 
formalism, we name $\cb_a$ the ``Blue's" function associated with $\gamn$. 
We now have to complete the arguments in the previous paragraph and show that
composition of such matrix valued functions of matricial arguments are
mathematically meaningful. 

In order that composition of such functions be a consistent operation (in the sense that it generalizes composition of complex-analytic functions of a 
complex variable), it is necessary that these matrix valued functions (with $\gmn$ in (\ref{gmn}) being an important example), have the same matricial texture as $\invp$ (Eq. (\ref{inverseprop})). Recall that (Eq. (\ref{gmn}))
\beqast
\gmn (\eta; z, z^*) = {1\over r^2 - (\eta-\Si)^2} ~
\left(\begin{array}{cc} \Si-\eta ~~~ & z\\{} & {}\\
z^* & \Si-\eta\end{array}\right)\,,
\eeqast
where $\eta$ and $\Si\equiv\Si(\eta;r)$ are complex. Thus, for $\eta$ complex, the texture of $\gmn$ is more complicated than the texture of
\beqast
\cg_0^{-1} =\left(\begin{array}{cc} \eta~~~ & z\\{} & {}\\
z^* & \eta\end{array}\right)
\eeqast
(Eq. (\ref{inverseprop}), whose off-diagonal blocks are hermitean conjugates of each other. This mismatch of textures can be dealt with readily by 
generalizing the space of matrices out of which $\invp$ is taken into 
\beq\label{invpropnew}
\cg_0^{-1} =\left(\begin{array}{cc} \eta~~~ & z\\{} & {}\\
w & \eta\end{array}\right)\,.
\eeq
Here the off diagonal blocks are no longer related. To see that this generalization is the natural one to make, we follow the first few steps of analysis in Section 4 and find that with (\ref{invpropnew}) the self-energy matrix is 
\beq\label{Sigmamnnew}
\Sigma^\mu_\nu = \Si(\eta; zw) \delta^\mu_\nu\,,
\eeq
(where the only change in (\ref{Sigmamn}) is that now $\Si$ depends on $zw$
rather than on $r^2=z^*z$.) Thus, from the definition of self-energy (\ref{selfenergy}) we immediately see that (\ref{gmn}) is changed into 
\beq\label{gmnnew}
\gmn (\eta; z, w) = {1\over zw - (\eta-\Si)^2} ~
\left(\begin{array}{cc} \Si-\eta ~~~ & z\\{} & {}\\
w & \Si-\eta\end{array}\right)\equiv \left(\begin{array}{cc} \cg ~~~ & G\\{} & {}\\ G'~~~~ & \cg\end{array}\right)\,.
\eeq
This has essentialy the same texture as (\ref{gmn}). With this extension, we have gained the fact that (\ref{invpropnew}), which generalizes here the ordinary complex variable $z$, has the same matricial texture as (\ref{gmnnew}).
At the end of our calculations we will set $w=z^*$. We can also see that the new texture (\ref{invpropnew}) is the right one at a more elementary level, by simply recognizing that the texture
of (\ref{invpropnew}) is preserved by raising this new $\invp$ to any integer (positive or negative) power.

With this texture it is enough to consider $\cg_a$ and $\invp$ simply as $2\times 2$ complex matrices with equal diagonal elements. Viewing $\cg_a\,\,(a=1,2)$ as a $2\times 2$ matrix valued function of a matricial argument (\ref{invpropnew}) is an equivalent way of saying that we are dealing with the three functions  $\cg_a(\eta; zw), G_a(\eta; z, w)$ and $G_a'(\eta; z, w)$, and composition of such matrix valued functions amounts
to compositions of such triads in the obvious way.

In the following the symbols $\cb, \cw, \cx$ etc. will stand for matrices of 
a texture identical to that of $\invp$ and $\cg_a$. The matrix valued Blue's function $\cb_a\left(\cw\right)$, associated with the probability distribution $P_a(\phi_a^\dgg\phi_a),\,(a=1,2)$ is defined then as the matrix 
\beq\label{B}
\cb_a =\left(\begin{array}{cc} B_a~~~ & b_a\\{} & {}\\
b_a' & B_a\end{array}\right)\,,
\eeq
that satisfies
\beq\label{blues}
\cb_a\left[\cg_a\left(\cx\right)\right] = \cx\,.
\eeq
Similarly, we define the $2\times 2$ matrix ``Blue's" function $\cb$ associated with (\ref{prob12}) by $\cb\left[\cg\left(\cx\right)\right] = \cx$.
From the definition of self-energy (\ref{selfenergy}), or equivalently from (\ref{selfenergya}), we readily identify
\beq\label{explicitB}
\cb(\cw) = \cw^{-1} + {\bf 1}\Si(\cw) = \cw^{-1} + {\bf 1}\sum_{k=1}^\infty \Gam_{2k} \left({1\over 2}\rmtr\cw\right)^{2k-1}\,,
\eeq
where ${\bf 1}$ is the $2\times 2$ unit matrix.

The identity of textures on both sides of (\ref{blues}) is assured already by the definitions of the quantities involved, and does not pose by itself any constraints on the various matrix elements that appear on both sides of (\ref{blues}).  In other words, $\cb_a$ is indeed the matrix valued inverse function of the 
matrix valued function $\cg_a$. Furthermore, due to this equality of textures 
we can readily apply $\cg_a$ to both sides of (\ref{blues}) obtaining 
\beq\label{blues1}
\cg_a\left[\cb_a\left[\cg_a\left(\cx\right)\right]\right] = \cg_a\left(\cx\right)\,,
\eeq
which verifies that the matricial function composition in (\ref{blues}) 
respects the obvious condition that taking the functional inverse twice 
in a row yields the original function. Thus, setting for example $\cx=\invp$ in (\ref{blues}), that equation simply says that 
starting with the set of three functions $\cg_a(\eta; zw), G_a(\eta; z, w)$ and $G_a' (\eta; z, w)$, we construct the set of three functions $B_a(\eta; zw), b_a(\eta; z, w)$ and $b_a'(\eta; z, w)$, such that $\cg_a(B_a; b_a' b_a)=\eta, G_a(B_a; b_a, b_a')=z$ and $G_a' (B_a; b_a, b_a')=w$. 

Note from (\ref{explicitB}) that $\cb(-\cw)=-\cb(\cw)$. This is so because 
$\cb$ is the functional inverse of the propagator $\cg(\cx)$ 
which is manifestly an odd function of $\cx$ due to the fact that in this paper all the probability distributions are even in $\phi$.

Let us now go back to (\ref{Sigmaa}). The propagator $\gamn$ on the right hand side of (\ref{Sigmaa}) is a function of $\eta, z$ and $w$, namely, of the
matrix $\invp$. We may generalize and consider it a function of the variable $\cw$, rendering $\sia$ a function of $\cw$, namely, $\sia (\cw) = \sum_{k=1}^{\infty} \Gak \left[{1\over 2}~\rmtr~\ga (\cw)\right]^{2k-1}$ as in (\ref{explicitB}). Similarly, by replacing the argument $\invp$ of $\gmn$ on the right hand side of (\ref{Sigmatot}) by a variable $\cx$ (with $\cx$ having the same texture as all the matrices above), we may consider the function $\Si(\cx) =
\sum_{k=1}^{\infty} \Gak \left[{1\over 2}~\rmtr~\cg (\cx)\right]^{2k-1}$.

Evaluating $\sia (\cw)$ at $\cw=\cw_a\equiv\cb_a \left[\cg \left(\invp\right)\right]$, 
and using (\ref{blues}) and (\ref{Sigmatot}) we immediately find that 
\beq\label{bluea}
\sum_a~\sia \left(\cw_a\right) = \sum_{k=1}^{\infty} \left(\sum_a\Gak\right) 
\left[\cg (\eta; r)\right]^{2k-1}\equiv \Si (\eta; r)\,.
\eeq
Thus, from (\ref{explicitB}) and from the matrix inverse of (\ref{blues}) we obtain 
\beqast
\sum_{a=1,2}\left\{\cb_a\left[\cg\left(\cx\right)\right]-\cg^{-1}\left(\cx\right)\right\}=\cx-\cg^{-1}\left(\cx\right)\,,
\eeqast
namely, 
\beq
\sum_{a=1,2}\cb_a\left[\cg\left(\cx\right)\right] - \cg^{-1}\left(\cx\right) = \cx\,.
\eeq
Therefore, by definition, the Blue's function associated with (\ref{prob12})
is simply
\beq\label{addition}
\cb_{T}\left(\cw\right) = \cb_{1}\left(\cw\right) + \cb_{2}\left(\cw\right)
-\cw^{-1}\,,
\eeq
in agreement with \cite{stony}. 

To summarize, we carried out diagrammatic expansions (namely, expansions in traces of powers of $\gamn$ and $\gmn$) of the self-energies
$\sia$ and $\Si$. We were allowed to do so because by construction $\invp$ was taken to be non-hermitean, and thus in the analyticity domain of $\gmn$.
Those diagrammatic expansions led us to introduce the matrix valued Blue's functions (\ref{B}), in close analogy with the addition formalism for ordinary hermitean random matrices \cite{bluez}. As in the latter case, the Blue's functions follow a simple addition formula (\ref{addition}) under 
multiplication of the individual probability distributions (\ref{prob12}). 

Now that we have $\cb_T$ in (\ref{addition}), we can calculate the Green's 
function $\gmn (\eta; z, w)$ associated with $H_T$ by evaluating the functional inverse of $\cb_T$ in (\ref{addition}) at argument $\invp$. After completing this step we may return to the original conventions and set $w=z^*$. 
Also, for the sole purpose of determining the eigenvalue distribution of $\phi=\phi_1+\phi_2$ all we need is $\gmn(0-i0;z, z^*)$, and we may also 
set $\eta=0-i0$.

\subsection{A Simple Application: A Non-abelian Non-Hermitean Central Limit Theorem}

Gauss proved that if we add $K$ random numbers $x_i$,
$i=1,2,...K$,
with $x_i$ taken from the probability distribution $P_i(x_i)$,
then the
normalized sum $s= {1\over \sqrt{K}} \sum_i x_i$ follows the
Gaussian
distribution in the limit $K$ tending to infinity. This result
plays an
important role in physics and mathematics and accounts for
the ubiquitous
appearance of the Gaussian distribution.

A generalization of Gauss's central limit theorem for the case in which the variables $x_i$ are replaced by $N\times N$ hermitean random matrices
$\varphi_i$, $i=1,2,...K$, taken from from the probability distributions
\beq
P_i(\varphi_i)={1 \over Z_i}{e^{{-N}
trV_i(\varphi_i)}}.
\label{distributioni}
\eeq
certainly exists and was recently discussed in \cite{bluez}.\footnote{Note that there exists another commonly considered class of random matrices, in which the element of the random matrices is each taken from a probability distribution (the same for each element). For this class, which is sometimes referred to as the Wigner distribution\cite{french}, the theorem follows immediately from
the usual abelian central limit theorem. Here we are speaking of the trace
classes defined by (\ref{distributioni}).} In this version 
of the theorem the normalized sum $\varphi_T \equiv {1\over\sqrt{K}} \sum_i \varphi_i$ indeed follows the Gaussian distribution in the large $K$
limit.

In this sub-section we wish to discuss Gauss's theorem for $N\times N$ non-hermitean random matrices
$\phi_i$, $i=1,2,...K$, that are taken from from the probability distributions
\beq
P_i(\phi_i^\dgg\phi_i)={1 \over Z_i}{e^{{-N}
trV_i(\phi_i^\dgg\phi_i)}}.
\label{distributionii}
\eeq
As was already remarked in\cite{bluez}, it is not difficult to generalize one 
of the standard proofs of Gauss's theorem to matrices, and we shall do so at the end of this subsection. However, we find it quite interesting 
to address the question of whether $\phi_T \equiv{1\over\sqrt{K}} \sum_i\phi_i$ follows the Gaussian distribution by applying the non-hermitean addition formalism just presented. In doing so, we shall generalize the analysis made in the last section of \cite{bluez} to non-hermitean random matrices and show that Ginibre's disk distribution \cite{ginibre} for the eigenvalues of $\phi_T$ emerges naturally from the addition formalism.

It is clear from the derivation of the addition formalism just given 
that when we add two random matrices, in general it is
difficult to determine explicitly the resulting $\gmn$ for the
sum of the two matrices. What we hope for here is that the large
$K$ limit will bring considerable simplification. This is indeed the
case.

From the law of 
addition (\ref{addition}) 
we learn that the
function $\cb(\cw)$ associated with the unknown $\cg(\cx)$ is given
by
\beq
\cb(\cw)=\sum_{i=1}^K\cb_a(\cw) - {K-1\over \cw}\,.
\label{sumB22}
\eeq
Using (\ref{explicitB}) we may rewrite (\ref{sumB22})
as
\beq\label{sumB222}
\cb(\cw)=\sum_{k=1}^\infty\left(\sum_{i=1}^K\Gam_{2k}^{(i)}\right)\left({1\over 2}\rmtr\cw\right)^{2k-1} + {1\over \cw}\,,
\eeq
where $\Gam_{2k}^{(i)}$ is the 2k-th cumulant associated with the distribution
$P_i$ in (\ref{distributionii}). 

By definition (Eq. (\ref{B})), the Green's function $\cg(\cx)$ associated with $\sum_i\phi_i$ is determined by solving $\cb(\cg(\cx))=\cx$.
 We are however interested in the normalized sum
$\phi_T \equiv {1\over \sqrt{K}}\sum_i \phi_i$. Define the corresponding propagator as
\beq\label{cgt}
\cg_T(\cx)\equiv\langle\left({1\over \cx- {1\over \sqrt K}\sum_i\ch_i}\right)\rangle=\sqrt{K}\cg({\sqrt K}\cx)\,,
\eeq
where
\beqast
\ch_i=\left(\begin{array}{cc} 0~~~ & \phi_i\\{} & {}\\
\phi^\dgg_i & 0\end{array}\right)\,.
\eeqast
Rescaling and substituting $\cg(\sqrt{K}\cx)$ into (\ref{sumB222}), we find 
that $\cg_T(\cx) $ is determined by
\beq
{1\over \cg_T(\cx)}+\sum_{k=1}^\infty\left[{1\over K^k}\left(\sum_{i=1}^K \Gam_{2k}^{(i)}\right)\left({1\over 2}\rmtr\cg_T(\cx)\right)^{2k-1}\right] =\cx\,.
\label{twidd}
\eeq

For finite $K$, (\ref{twidd}) is hopelessly complicated. However, as $K\rightarrow\infty$ we see that it simplifies rather naturally to
\beq
\cg_T^{-1}(\cx)+{\Gam_2\over 2}~\rmtr\cg_T(\cx)=\cx
\label{twidd2}
\eeq
with
\beq
\Gam_2 \equiv {1\over K} \sum_i^K \Gam_2^{(i)}\,.
\label{gam2}
\eeq
The self-energy associated with (\ref{twidd2}) is evidently
$\Si_T = {\Gam_2\over 2}~\rmtr\cg_T(\cx)$. It is the simplest case of 
the general formula (Eq.(\ref{Sigmaa})) $\Si(\cg) = \sum_{k=1}^{\infty}\Gam_{2k}\left({1\over 2}~\rmtr\cg\right)^{2k-1}$, 
where only $\Gam_2\neq 0$. In other words, it is equal to the self-energy
associated with the Gaussian distribution 
\beq\label{gausseffective}
P_G(\pdgp) = {1\over Z} {\rm exp} \left(-{N\over \Gam_2}~\rmtr \pdgp\right)\,.
\eeq
Re-substituting
\beqast
\cx=\invp = \left(\begin{array}{cc} \eta~~~ & z\\{} & {}\\
z^* & \eta\end{array}\right)\,,
\eeqast
and using the general hermitization formalism in Section 2 (and in particular Eqs. (\ref{rho}) and (\ref{greens1})) we conclude \cite{stony, fz} that the eigenvalues of $\phi_T = (1/\sqrt{K})\sum_i^K\phi_i$ are distributed uniformly in a disk of radius $\sqrt{\Gam_2}$.

Thus, not only do we obtain Ginibre's distribution, we also learn that $\Gam_2$ is given by (\ref{gam2}).

While we have proved that the density of eigenvalues of
$\phi_T$ is identical to the density associated with the Gaussian distribution (\ref{gausseffective}), we cannot yet conclude that the probability
distribution of $\phi_T$ is Gaussian. The reason is that in principle, a given eigenvalue distribution may correspond to more than a single probability distribution\footnote{This is indeed the case for hermitean random matrices. 
For example, for matrices in the Wigner class, even when the
distribution is not Gaussian, the corresponding density of eigenvalues still satisfies the semi-circle law, as is well known.}. To see that $\phi_T$ is indeed taken from a Gaussian distribution, we have to calculate that distribution explicitly. 

As mentioned above, it is not difficult to extend one of the
usual proofs of the central limit theorem to the case of matrices. The distribution for the normalized sum matrix
$\phi \equiv {1\over \sqrt K}\sum_i \phi_i$ is given
by (here we omit the subscript $T$)
\beqra
P(\phi) &=& \left({\prod_i^K} \int d\phi_i P_i(\phi_i^\dgg\phi_i)\right)
\delta\left(\phi - {1\over
\sqrt K}\sum_i \phi_i\right)\nonumber\\{}\nonumber\\
&=&\int dt \left( {\prod_i^K} \int d\phi_i P(\phi_i^\dgg\phi_i)\right) 
e^{ {i\over\sqrt K}\sum_i^K ~\Im\rmtr ~t \phi_i} e^{-i ~\Im\rmtr ~t\phi}\,.
\label{sumdist}
\eeqra
The integral over the $\phi_i$ can be done in the large $K$
limit:
\beqra
&&\int d\phi_i P(\phi_i) e^{ {i\over \sqrt K}\sum_i ~\Im\rmtr ~t\phi_i }\nonumber\\
&=& 1 - {1\over 4K}\int d\phi_i P(\phi_i^\dgg\phi_i) ~\rmtr~(t\phi_i)~\rmtr
(t^\dgg\phi_i^\dgg) + {\cal O}\left( {1\over K^2}\right)\nonumber\\
&=& 1 - {\Gam_2^{(i)} \over 4KN} ~\rmtr~(t^\dgg t)  +
{\cal O}\left( {1\over K^2}\right)
\label{quad}
\eeqra
where by definition $\Gam_2^{(i)} = \int d\phi_i
P(\phi_i^\dgg\phi_i) {1\over
N} ~\rmtr
\phi_i^\dgg\phi_i$.  Reexponentiating and integrating over $t$ we
obtain the desired result that $P(\phi)$ is proportional to 
$e^{- {N\over \Gam_2} ~\rmtr\pdgp }$, with $\Gam_2$ given by (\ref{gam2}).

%************************************************************
%P.4

\vskip 20mm
\begin{center}
{\bf ACKNOWLEDGEMENTS}
\end{center}
A.Z. would like to thank the Institute for Advanced Study for a Dyson Distinguished Visiting Professorship and also Freeman Dyson for helpful conversations. This work was supported in part by the 
National Science Foundation under Grant No. PHY89-04035, and by the Dyson Visiting Professor Funds. Both authors thank R. Scalettar for performing
numerical checks of this work, which uncovered an error in Section 5.1.

\pagebreak
%******************************************
%   Appendix 
%************************************************************

\newpage
\setcounter{equation}{0}
\renewcommand{\theequation}{A.\arabic{equation}}
{\bf Appendix : {A Note Concerning the ``Single Ring Theorem"}}
\vskip 5mm

The hermitean matrix $\pdgp$ can always be diagonalized $\pdgp=U^\dgg\La^2 U$ by a unitary matrix $U$, with $\La^2={\rm diag}(\la_1^2, \la_2^2,\cdots,\la_N^2),$ where the $\la_i$ are all real. This implies that $\phi=V^\dgg\La U$, with $V$ a unitary matrix as well. Thus, the complex eigenvalues of $\phi$ are given by the roots of ${\rm det} (z-\La W)=0$, with $W=UV^\dgg$. Evidently, as $W$ ranges over $U(N)$, the eigenvalues of $\La W$ could be smeared (in the sense that 
they would not span narrow annuli around the circles $|z|=|\la_i|$.)
In Section 6, we proved (in the large $N$ limit) that for $\phi$ taken from 
the class of ensembles of the form $P(\pdgp)=(1/Z) {\rm exp}[-N\rmtr~V(\pdgp)]$ (Eq. (\ref{prob})), the eigenvalues can only span either a disk or an annulus.We have verified that this is the case, explicitly for $N=2$, and numerically for large $N$ (of the order of $100$) and for some range of $\La^2$.

\end{document}